\documentclass[twocolumn,amsmath,superscriptaddress,amsfonts,prb]{revtex4}
\usepackage{graphicx}
\usepackage{subfigure}
\begin{document}
\title{Dimensional quantization effects in the thermodynamic of conductive filaments}
\author{D. Niraula}\email{dipesh.niraula@rockets.utodeo.edu}
\author{C. R. Grice}\email{corey.grice@rockets.utoledo.edu}
\author{V. G. Karpov}\email{victor.karpov@utoledo.edu}\affiliation{Department of Physics and Astronomy, University of Toledo, Toledo, OH 43606, USA}
\begin{abstract}
We consider the physical effects of dimensional quantization in conductive filaments that underlie operations of some modern electronic devices. We show that, as a result of quantization, a sufficiently constricted filament acquires a positive charge. Several applications of this finding include describing the local host material polarization, the stability limit of filament constrictions, equilibrium filament radius, effects of polarity in device switching, and quantization of conductance.
\end{abstract}
\maketitle

The role of conductive filaments (CFs) is critically important for functionality of such electronic devices as phase change memory (PCM), \cite{PCM} resistive random access memory (RRAM) \cite{RRAM}, and related threshold switches (TS); CFs are responsible for the breakdown phenomena in gate dielectrics and some other structures. It is customary to describe CFs as thin metallic cylinders with appropriate material parameters. That approach requires modifications for modern devices where CF radii ($R$) fall in the nanoscale domain, thus bringing up the phenomenon of dimensional quantization.

\begin{figure}[b!]
\includegraphics[width=0.33\textwidth]{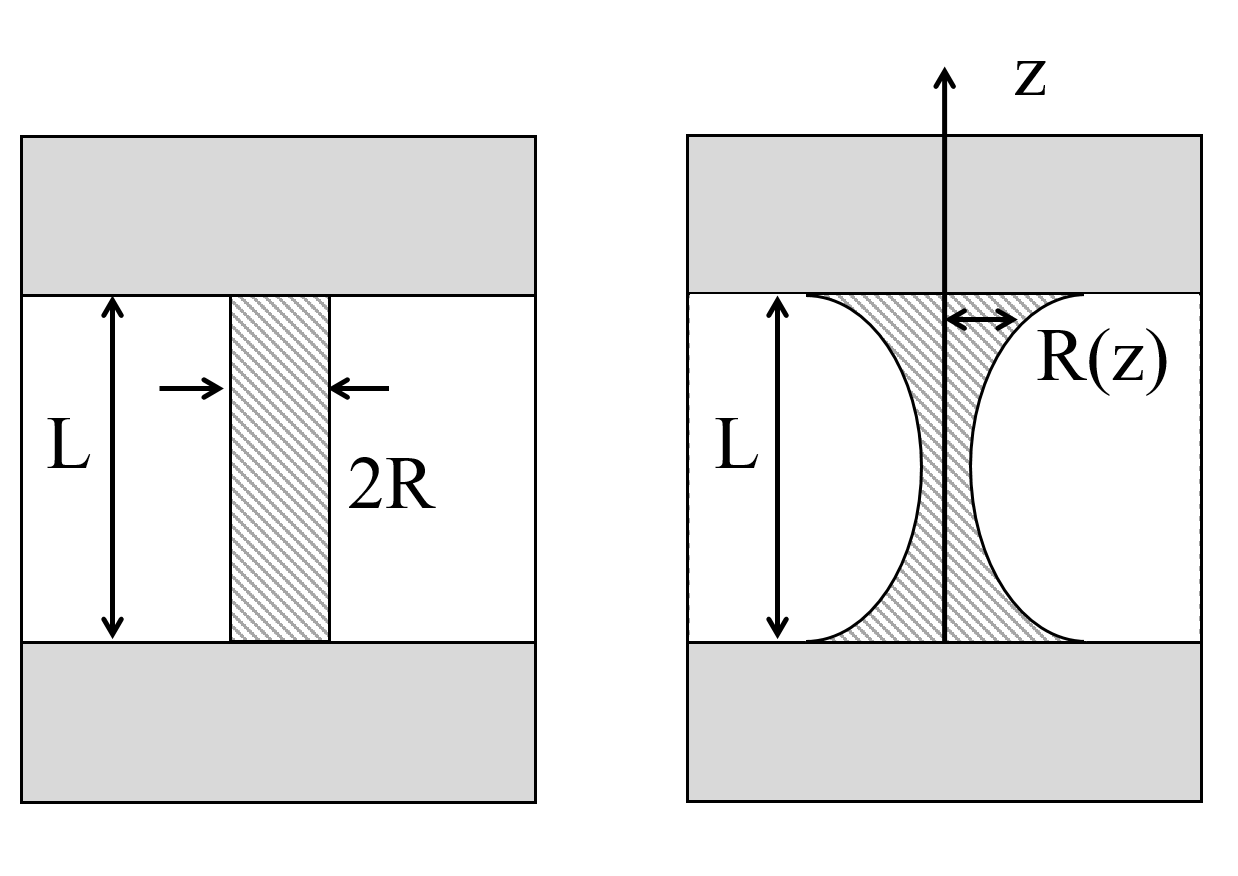}
    \caption{A uniform (left) and constricted (right) CF.}\label{Fig:CF}
    \end{figure}

Here we consider mostly the effects of dimensional quantization in CF thermodynamics. We show that such effects are responsible for CF equilibrium radius, nucleation barrier, and the instability of CF constrictions; in addition, we briefly revisit the concept of quantum point contacts (QPC) developed for constricted CFs. \cite{degrave2010,lian2014,miranda2008,li2015} Schematic representations of the types of objects dealt with here are illustrated in Fig. \ref{Fig:CF}.

\begin{figure}[b!]
\includegraphics[width=0.5\textwidth]{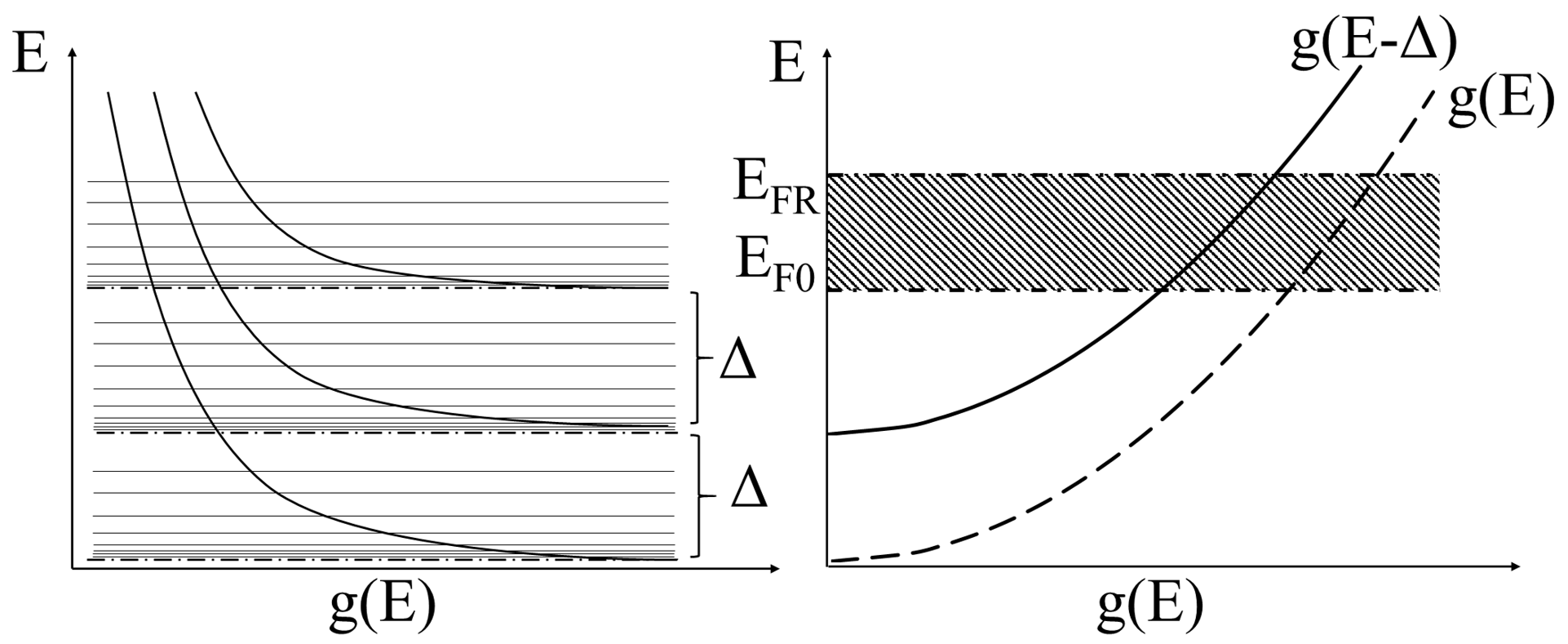}
    \caption{Left: 1D sub-bands density of states along z-direction separated by energy $\Delta$ due to the quantum confinement in a circular potential box  of radius R. Right: The density of states of a CF with and without dimensional quantization. $E_{F0}$ stands for the reservoir Fermi energy; $E_{FR}$ is the Fermi energy in a stand alone CF of radius $R$.}\label{Fig:1DDos}
    \end{figure}

First, we consider a uniform CF as a long ($L\gg R$) metal cylinder embedded in a dielectric. As described in cylindrical coordinates, the 2D quantization in the radial direction produces discrete energy levels $E_n,\quad n=1,2,...$, while the much longer length $L$ in $z$-direction corresponds to sub-bands of practically continuous one-dimensional density of states, $g_1(E-E_n)\propto 1/\sqrt{E-E_n}$ above each of $E_n$. The latter form spikes in the density of states at $E_n$ as illustrated in Fig. \ref{Fig:1DDos}.

The sub-band overlap will change the density of states at the Fermi level, thus affecting the system thermodynamics. More quantitatively, $E_n$ represents the n-th energy level of a circular potential box of CF radius $R$, and values of $E_n$ are determined by the zeros of the Bessel function. \cite{landau1991} For the highly excited states, the problem is simplified by the relation \cite{robinett2003} between the number of energy levels $n(E)$ below $E$ in a box of area $A$ and perimeter $P$,
\begin{equation}\label{Eq:Nooflevel}
n(E) = \frac{A}{4\pi}\frac{2m}{\hbar ^2}E - \frac{P}{2\pi}\sqrt{\frac{2mE}{\hbar^2}}.
\end{equation} with $m$ being the electron mass.
Setting $A=\pi R^2$ and $P= 2\pi R$, we obtain, $dn/dE$, the separation between energy levels $\Delta=(dn/dE)^{-1}$, and the spectrum,
\begin{equation}\label{eq:En}E_n=n\Delta , \quad \Delta=\frac{2\hbar ^2}{mR^2} \quad {\rm when}\quad n\gg 1\end{equation}
where we have neglected the second term in Eq. (\ref{Eq:Nooflevel}) for the highly excited states, $n\gg 1$. For $R\sim 1$ nm, $\Delta\sim 0.1$ eV and is much smaller than the Fermi energy $E_F\sim 3-10$ eV, which justifies the excited states approximation. As usual, $E_F$ is assumed to be fixed by the system electronic reservoir and does not depend on the CF characteristics.

The density of states in the CF $g(E)$ is calculated by summation of the 1D sub-band  contributions $g_1(E-E_n)\Theta(E-E_n)$ for all the energy levels from 1 to $E_n=E$ and dividing the result by the filament volume $\pi R^2L$. Here, $\Theta (x)=\{1 \ {\rm when}\ x \geq 0;\ 0 \ {\rm when}\ x < 0\}$ and  $g_1(E) = (L/\hbar\pi )\sqrt{m/2E}$. Replacing the summation with integration yields the envelope density of states,
\begin{multline}\label{Eq:summation}
g(E)=\frac{1}{\pi R^2}\sum_{n=1}^{E/\Delta} \frac{\Theta(E-E_n)}{\sqrt{2}\pi\hbar}\sqrt{\frac{m}{E-E_n}} \simeq \\
\int\limits_{1}^{E/\Delta}\frac{\Theta(E-E_n)\sqrt{m}dn}{\pi ^2R^2\hbar\sqrt{2(E-E_n})}=  \frac{\sqrt{2}m^{3/2}\sqrt{E-\Delta }}{\pi^2 \hbar^3 }.
\end{multline}

The latter result neglects $g_1$ spikes illustrated in Fig. \ref{Fig:1DDos} compared to the smooth envelope obtained by integration of many $g_1(E)$ tails. That approximation is justified when a spike value $g_1(E-E_n)$ corresponding to a thermal energy $E-E_n\sim kT$, is smaller than that of the envelope in Eq. (\ref{Eq:summation}), which reduces to the inequality,
\begin{equation}\label{eq:kT}
kT\gg \frac{\pi}{16}\frac{\Delta ^2}{E-\Delta }.\end{equation}
Since $E-\Delta\approx E_F$, it is well obeyed for practical temperatures, say $T\gtrsim 300$ K.


Eq. (\ref{Eq:summation}) naturally predicts the standard 3D density of states $g_3(E)\propto \sqrt{E}$ when the CF radius is large, i. e. $\Delta\rightarrow 0$. For finite $\Delta$, it follows from Eq. (\ref{Eq:summation}) that dimensional quantization shifts the standard 3D energy spectrum up by energy $\Delta$, as illustrated in Fig. \ref{Fig:1DDos} (right). That shift makes electron energy levels in a band $\Delta$ be raised in energy above $E_F$. Such levels will thus lose their electrons to the lower energy states outside the CF, rendering it with a net positive charge.

\begin{figure}[b!]
\includegraphics[width=0.3\textwidth]{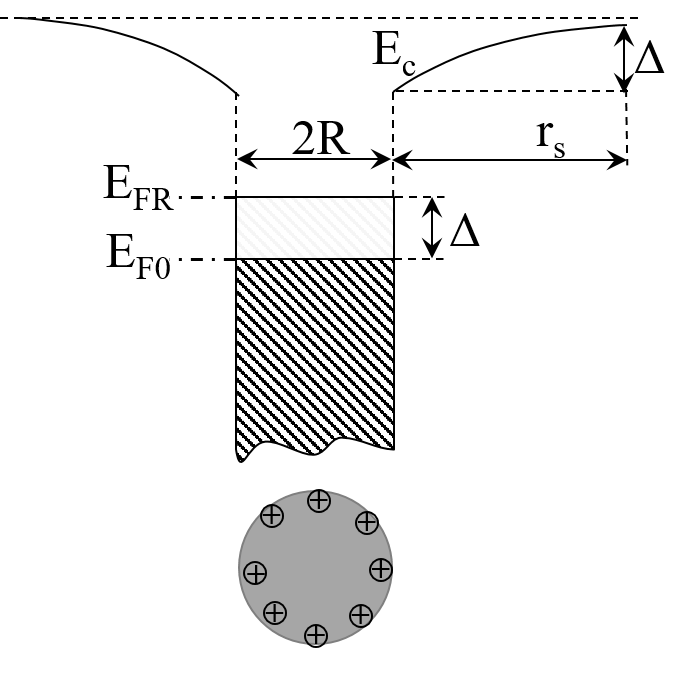}
    \caption{Top: The energy band cross section of CF and its surrounding host. The difference $\Delta =E_{FR}-E_{F0}$ is balanced by the band bending shown for the conduction band edge $E_c$ induced by CF charging. Bottom: a real space filament cross section with surface positive charges.}\label{fig:band}
    \end{figure}

An alternative wording is that the upward shift of energy spectrum increases the CF Fermi energy by $\Delta$. To restore the equilibrium with the host material Fermi energy, the CF gives away its electrons acquiring a positive charge as illustrated in Fig. \ref{fig:band}. The net charge of the CF must be confined to its (sub-nanometer thin) surface layer as always for metals. The inner region of the CF remains intact with electron energies shifted down due to the electrostatic potential of surface charge.

Denoting by $\Lambda$ the CF's linear charge density, the electric field strength outside the CF becomes ${\cal E}=2\Lambda/\kappa r$ at distances $r$ smaller than the screening length, $r_s$ where $\kappa$ is the dielectric permittivity. We approximately account for the screening by cutting off the electric potential perturbation at $r=r_s$, which provides the downward electron energy shift $(2\Lambda e/\kappa )\ln (r_s/R)$. Setting the latter equal to $\Delta$ yields,
\begin{equation}\label{eq:Lambda}\Lambda=\frac{\kappa\Delta}{2e\ln(r_s/R)}.\end{equation}

We assume a total of at least several electrons are present within the CF, so charge quantization is not important. Our numerical estimates utilize the parameters characteristic of CFs in HfO$_2$ based RRAM devices: $R\sim 1$ nm \cite{celano2014,privetera2013} and $\kappa \sim 25$, the effective electron mass \cite{monaghana2009} $m\sim (0.1-1)m_0$ where $m_0$ is the true electron mass. The screening radius $r_s\sim L\sim 10$ nm is always limited to the distance between the metal electrodes due to their electron redistribution. \cite{cooray2007} With these parameters, Eq. (\ref{eq:Lambda}) yields the number of electrons in a single CF, $\Lambda /e\sim 5-50$.

One additional condition of CF charging is that its related decrease in electronic energy,
\begin{equation}\label{eq:FDelta}F_{\Delta}=g(E_F)(\Delta ^2/2)\pi R^2L\end{equation}
exceeds the electric field energy
\begin{equation}\label{eq:loss}F_{\cal E}=\int \frac{\kappa {\cal E}^2}{8\pi}d^3r=\frac{\Delta ^2\kappa L}{4e^2\ln(r_s/R)}\end{equation} where we have used $\Lambda$ from Eq. (\ref{eq:Lambda}) and treated the logarithm in the integrand as a constant. The resulting inequality,
\begin{equation}\label{eq:gain}2\pi g(E_F)(e^2/\kappa)R^2\ln (r_s/R) >1\end{equation}
obeys for all practical parameters.

Note that CFs can be charged regardless of dimensional quantization due to the difference $\delta E_F$ between CF and the host material Fermi energies. That additional linear charge density, $\Lambda '$, related to the CF electric capacitance \cite{karpov2017} is given by Eq. (\ref{eq:Lambda}) with $\delta E_F$ instead of $\kappa\Delta$. While typically insignificant in high-$\kappa$ materials, $\Lambda '$ can be easily accounted for by redefining $\Lambda \rightarrow \Lambda +\Lambda '$.

Consider several applications of the above theory.

{\it CF field, polarization, and instability.} For a 1 nm radius CF, the lateral field strength at its surface is estimated as ${\cal E}\sim 10$ MV/cm. It can significantly polarize the host material by aligning ferroelectric domains (found e. g. in doped HfO$_2$ films \cite{fa2016}) or moving charged defects. Such a polarization can serve as a readable nonvolatile memory. On the other hand, according to Eq. (\ref{eq:loss}) the field strength and energy increase, respectively, as $R^{-2}$ and $R^{-4}$ with filament thinning. That points at the fundamental instability of CF constrictions in which they will evolve (breaking or healing) to decrease the system energy.

{\it Equilibrium radius.} The field energy of Eq. (\ref{eq:loss}) favors CF radius increase, while its surface energy $2\pi RL\sigma$ has the opposite trend, where $\sigma$ is the interfacial energy.
The sum of these two contributions in CF free energy $F$ has a minimum, $dF/dR=0$, that determines the equilibrium radius $R_0$ of CF through the relation,
\begin{equation}\label{eq:R0}
R_0=\left[\frac{2\hbar^4\kappa}{\pi e^2m^2\sigma\ln(r_s/R_0)}\right]^{1/5}
\end{equation}
Calculations made using the typical \cite{jeurgens2009} $\sigma\sim 100$ dyn/cm and other relevant parameters mentioned above yields $R_0\sim 0.5-1.5$ nm, which is consistent with many of the observed CF radii. \cite{celano2014,privetera2013} Note that $F_{\Delta}$ does not contribute to the above free energy because of the cancelation with the increase of electron energies illustrated in Figs. \ref{Fig:1DDos} (right) and \ref{fig:band}. The contribution from the CF vs. host bulk material transformation is estimated to have a minor effect for their typical chemical potential differentials.
\begin{figure}[t!]
\includegraphics[width=0.48\textwidth]{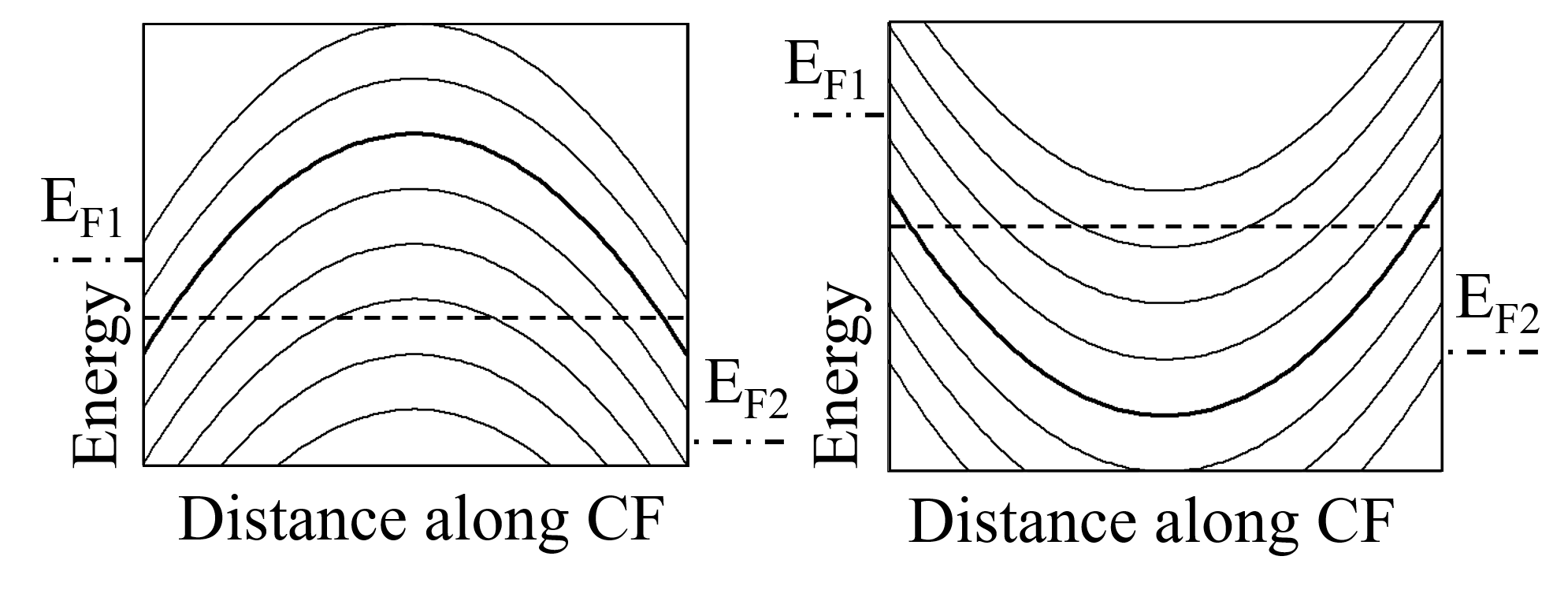}
    \caption{{\it Left}: negative curvature barriers corresponding to the constriction dominated channels (between two consecutive parabolas each). As an illustration, the electron with energy depicted by a horizontal dash line must overcome the barrier formed by the intersection of that line and the bolded line parabola. {\it Right}: positive curvature conductive channels corresponding to a charged uniform CF.  Note two triangular barriers formed by the horizontal dashed line and the bolded line parabola. The region where that parabola falls below the dashed line is classically allowed.  }\label{fig:channels}
    \end{figure}

{\it Field induced nucleation.} Consider the nucleation of metallic needle-shaped particles in the electric field of high strength ${\cal E}_0$. The nucleus free energy is given by,
\begin{equation}\label{eq:FIN}
F=F_{\cal E}+2\pi RL\sigma -L^3\kappa {\cal E}_0^2
\end{equation}
where $F_{\cal E}$ is given by Eq. (\ref{eq:loss}) and the last term represents the polarization electrostatic energy gain written here to the accuracy of insignificant multipliers. \cite{karpov2008,karpov2008a} The term $F_{\cal E}$ distinguishes our analysis from the preceding work. The critical nucleus dimensions $R_0$ and $L_0$ are obtained from $dF/dR=dF/dL=0$. It is straightforward to see that $R_0$ is the same as in Eq. (\ref{eq:R0}), while $L_0$ and the nucleation barrier are a factor of $\sqrt{2}$ smaller than that of preceding theory.

{\it Bias polarity effects.} A new feature in field induced nucleation arises when we add the term, ${\cal E}_0\Lambda L^2/2$, for the energy of a linear charge $\Lambda$ in the external field. Because the sign of $\Lambda$ is predetermined, that term will make CF nucleation dependent on the field polarity. It was observed indeed that PCM operations are bias polarity dependent. \cite{lee2008,padilla2011} A theory here can be readily developed to describe that effect quantitatively.

We note another polarity effect for the bias induced CF charge. \cite{karpov2017} Adding it to $\Lambda$ from Eq. (\ref{eq:Lambda}) will increase or decrease the absolute value of the CF charge depending on the bias polarity; CF thinness will aggravate the asymmetry. That polarity effect can result in unequal absolute values of SET and RESET voltages for bipolar RRAM devices.

{\it Quantum conductance}.  In the QPC model, \cite{glazman1988} CF has a constriction; hence $R=R(z)$ as depicted in Fig. \ref{Fig:CF} and $\Delta (z)=2\hbar ^2/mR^2(z)$ adiabatically depends on the CF's longitudinal coordinate $z$. The transverse quantization is approximated by $E_n=n\Delta (0)-\alpha z^2$ presenting a set of parabolic barriers in Fig. \ref{fig:channels}(left) where $\Delta (0)$ and $\alpha$ are two constants. When the electric bias $V=[E_{F1}-E_{F2}]/e$ between the contacts with Fermi levels $E_{F1}$ and $E_{F2}$ changes, so does the number of conductive channels $N_{\rm ch}=[eV/\Delta (0)]$ where $[X]$ means the greatest integer in $X$. Assuming ballistic electron transport, the CF's conductance is given by $G_0N_{\rm ch}t$ where $G_0\equiv 2e^2/\pi\hbar \approx (12.9\ {\rm k}\Omega )^{-1}$ is the quantum conductance and $t$ is the barrier transparency due to  tunneling or activation. Recent work \cite{degrave2010,lian2014,miranda2008,li2015} attributed some RRAM observations to QPC model.

Our concept of positively charged CF, predicts electron energy decreasing towards the CF center regardless of constrictions. It is illustrated in Fig. \ref{fig:channels} (right) with positive curvature parabolas. Using standard electrostatics and $\Lambda$ from Eq. (\ref{eq:Lambda}), the curvature can be estimated as $K_{\rm PC}=4\kappa\hbar ^2/[mR^2L^2\ln(r_s/R)]$. This should be compared to the negative curvature of QPC parabolas estimated as $K_{\rm QPC}=-8\hbar ^2/(mR^2L^2)$ assuming that $R$ decreases over length $L/2$.

While the two are comparable in absolute value, their mutual cancelation would be a sheer coincidence. When the positive curvature dominates, the conductance quantization takes place as long  as the transport remains ballistic. Indeed, in spite of the energy modulation, the product of electron velocity $v=\sqrt{2E/m}$ times $g_1(E)\propto 1\sqrt{E}$ is energy independent, as required by the derivation. \cite{imry1997,datta2005}. If the transport is diffusive inside the classically allowed region of CF in Fig. \ref{fig:channels} (right), then two triangular barriers by the channel edges  exhibit additive QPCs. \cite{imry1997,datta2005}

The triangular barrier transparency $t$  is due to thermally assisted tunneling. \cite{karpus1986} Omitting the details, our analysis yields that such a CF conductance
\begin{equation}\label{eq:dG}
G=G_0N_{\rm ch}\exp\left[-\frac{\Delta}{kT}+\frac{\hbar ^2K_{\rm PC}^2L^2}{96m^2(kT)^3}\right]
\end{equation}
exponentially decreases with CF length and exponentially increases with its radius; non-Ohmicity can be introduced with non-ideal contacts. Eq. (\ref{eq:dG}) predicts the stepwise changes in conduction and shows that quantum conductance can be explained without the constriction hypothesis. [It is valid as long as the exponent is small, which otherwise should be approximated by unity.]

{\it Variability.} Dimensional quantization will contribute to variability of filamentary nano-devices, with amorphous structure. Eq. (\ref{Eq:Nooflevel}) predicts that random deviations from a perfect circular cross-section will create variability in the quantization energy $\Delta$ and its related effects. As an example , the number of conductive channels and the observed steps in conductance of Eq. (\ref{eq:dG}) will vary between different samples.\\

In conclusion, we have shown that dimensional quantization results in charging of conductive nano-filaments, which can exhibit itself in several effects of practical importance, such as strong polarization of the surrounding material, thermodynamic instability of constrictions, equilibrium filament radius, polarity effects in field induced nucleation, as well as quantization of a charged filament conductance.

This work was supported in part by the Semiconductor Research Corporation (SRC) under Contract No. 2016-LM-2654. We are grateful to I. V. Karpov and R. Kotlyar for useful discussions.

\end{document}